\documentclass[10pt]{article}

\usepackage[utf8]{inputenc}
\usepackage{amsmath}
\usepackage{amsfonts}
\usepackage{amssymb}
\usepackage{makeidx}
\usepackage{graphicx}
\usepackage{verbatim}
\usepackage{wrapfig}
\usepackage{xcolor}
\usepackage{multicol}
\usepackage{comment}
\usepackage{setspace}
\usepackage{soul}
\usepackage{makeidx}
\usepackage{braket}
\usepackage{multirow}
\usepackage[font=small,labelfont=it]{caption}
\usepackage[font=scriptsize]{caption}
\usepackage{blindtext}
\allowdisplaybreaks

\def\R{\mathbb R}

\newcommand{\si}{\sigma}

\newcommand{\cC}{{\cal C}}

\newcommand{\cH}{{\cal H}}

\newcommand{\+}{{\dagger}}
\newcommand{\sfrac}[2]{{\textstyle\frac{#1}{#2}}}
\newcommand{\half}{{\sfrac12}}
\newcommand{\pa}{\partial}
\newcommand{\res}{{\mathrm{res}}}
\newcommand{\weyl}{{\mathrm{weyl}}}
\newcommand{\Weyl}{{\mathrm{Weyl}}}

\newcommand{\im}{{\mathrm{i}}}

\newcommand{\ep}{{\mathrm{e}}}

\newcommand{\beq}{\begin{equation}}
\newcommand{\eeq}{\end{equation}}
\newcommand{\eq}{\end{equation}}
\newcommand{\bea}{\begin{eqnarray}}
\newcommand{\eea}{\end{eqnarray}}
\newcommand{\with}{{\quad{\rm with}\quad}}
\newcommand{\for}{{\quad{\rm for}\quad}}
\renewcommand{\and}{{\quad{\rm and}\quad}}
\newcommand{\und}{{\qquad{\rm and}\qquad}}
\renewcommand{\=}{\ =\ }

\advance \textheight by \topskip
\makeatletter
\@addtoreset{equation}{section}

\makeatother

\begin{document}

\begin{titlepage}
\setcounter{page}{0}

\phantom{.}
\vskip 1.5cm

\begin{center}

{\LARGE \bf 
A Casimir operator for a Calogero $W$ algebra 
}
\vspace{12mm}

{\Large Francisco Correa$^{a}$, Gonzalo Leal$^{b}$, Olaf Lechtenfeld$^{b\,\ast}$ and Ian Marquette$^{c}$}
\\[8mm]
\noindent ${}^a${\em 
Departamento de F\'isica\\
Universidad de Santiago de Chile,  Av.~Victor Jara 3493, Santiago, Chile}\\[6mm]
\noindent ${}^b${\em
Institut f\"ur Theoretische Physik and Riemann Center for Geometry and Physics\\
Leibniz Universit\"at Hannover, Appelstrasse 2, 30167 Hannover, Germany}\\[6mm]
\noindent ${}^c${\em
School of Mathematics and Physics\\
The University of Queensland, Brisbane, QLD 4072, Australia}
\vspace{12mm}

\begin{abstract} \noindent
We investigate the nonlinear algebra $W_3$ generated by the 9 functionally independent 
permutation-symmetric operators in the three-particle rational quantum Calogero model. 
Decoupling the center of mass, we pass to a smaller algebra $W'_3$ generated by 7 operators, 
which fall into a spin-$1$ and a spin-$\sfrac32$ representation of the conformal $sl(2)$ subalgebra. 
The commutators of the spin-$\sfrac32$ generators with each other are quadratic in the spin-$1$ 
generators, with a central term depending on the Calogero coupling. One expects this algebra to 
feature three Casimir operators, and we construct the lowest one explicitly in terms of Weyl-ordered 
products of the 7~generators. It is a polynomial of degree 6 in these generators, with coefficients 
being up to quartic in~$\hbar$ and quadratic polynomials in the Calogero coupling~$\hbar^2g(g{-}1)$. 
Putting back the center of mass, our Casimir operator for~$W_3$ is a degree-9 polynomial in the 
9 generators. The computations require the evaluation of nested Weyl orderings.  The classical and 
free-particle limits are also given. Our scheme can be extended to any finite number~$N$ of 
Calogero particles and the corresponding nonlinear algebras $W_N$ and $W'_N$.
\end{abstract}

\end{center}

\vfill

${}^\ast$\ Corresponding author email: olaf.lechtenfeld@itp.uni-hannover.de

\end{titlepage}

\section{Introduction and summary}

\noindent
The Calogero model~\cite{Cal,OlPe} is a workhorse of integrable systems.
It is known at least since 1995~\cite{Kuz,IsaLei} that the rational $N$-particle quantum Calogero model
(based on the $A_{N-1}\oplus A_1$ Coxeter system) features a $W_{1+\infty}$ algebra of permutation-invariant
observables, namely symmetric polynomials in the particle coordinates and Dunkl-deformed momenta. 
For a fixed number $N$ of particles, however, the polynomials of total degree larger than~$N$ 
can be expressed nonlinearly through polynomials of lower degree, which turns the linear
$W_{1+\infty}$ algebra into a nonlinear $W_N$ algebra~\cite{Poly2}. Furthermore, even the set of
polynomials of total degree up to~$N$ is overcomplete: their lower bound
$\half(N{+}1)(N{+}2)-1$ is larger than the $2N$ degrees of freedom due to coordinates and momenta.
Therefore, a number of classical identities relates them algebraically.
In the quantum theory, this redundancy translates to the existence of Casimir operators for
the $W_N$ algebra, one for each identity. Different representations of this quantum commutator algebra
(e.g.~in terms for partial differential operators) are obtained for different values of these Casimirs.

The two simplest invariants are the center-of-mass position and momentum of the $N$-particle system.
While classically they decouple and may be ignored, in the quantum theory they cannot be put to zero.
However, the $W_N$ algebra can still be decomposed into a center-of-mass part (a Heisenberg algebra)
and ``relative-motion part'', which we denote by $W'_N$. Once a suitable basis for $W'_N$ is found,
only invariants of total degree between 2 and~$N$ will occur. This does not change the expected lower bound
$\half N(N{-}1)$ of the number of Casimir operators.

For the two-particle case ($N{=}2$) the story is well known:
besides the center-of-mass position and momentum there are three quadratic invariants,
which close to an $sl(2)$ algebra. 
This (exceptionally linear) ``$W'_2$ algebra'' is the conformal algebra in $1{+}0$ dimensions 
and will occur as a subalgebra of all higher $W'_N$ algebras, 
which will help us to restrict the structure of the latter.
The single Casimir operator in this case is simply the $sl(2)$ Casimir~$\cC_2$, 
which is of second order in the $sl(2)$ generators and separately quadratic 
in the coordinates and Dunkl momenta, thus of total degree four.
Its eigenvalue, parametrized as $\hbar^2 g(g{-}1)$, becomes the Calogero coupling.

In this work we analyze the situation for three particles, $N{=}3$. 
After decoupling the center of mass, we are left with the three generators of $sl(2)$
plus four cubic invariants in a spin-$\frac32$ representation of $sl(2)$, 
which together form a $W'_3$ algebra. Nontrivially, two cubic invariants commute to
something quadratic in the $sl(2)$ generators, a term linear in~$\cC_2$, 
and a central extension (of relative order $\hbar^2$).
Only the latter feels the Calogero coupling.

We expect precisely three independent Casimir operators for this quadratic algebra.
Here, we identify $\cC_6$, which is the lowest one and of degree 6 separately in
the particle coordinates and Dunkl momenta, and also of order up to 6 in the $W'_3$ generators.
The quantum Casimir differs from the classical one by corrections of order $\hbar^2$ and $\hbar^4$,
and only the quantum corrections explicitly depend on the Calogero parameter~$g$.

The model (and the nonlinear algebra) simplifies in two different limits.
Sending the Calogero coupling to zero produces a system of three non-interacting particles.
Although the (permutation) invariant operators then reduce to homogeneous polynomials in 
coordinates and momenta, their algebra $W'_3\big|_{g=0}$ differs from $W'_3$ only in the central terms.
The latter are completely eliminated when $\hbar\to0$. However, the interacting classical limit is one
where the Calogero parameter becomes large, i.e.~$\hbar g=:\ell$ is fixed so that the strength of the Calogero
potential becomes~$\ell^2$. In this limit a complementary part of the central terms is removed,
leading to $W'_3\big|_{\textrm{class}}$. 

Casimir operators are used to classify irreducible representations. For $sl(2)$,
the well-known unitary continuous representations are infinite-dimensional with $\cC_2\ge-\sfrac14\hbar^2$,
thus parametrized by~$g\ge\sfrac12$. For our nonlinear algebra~$W'_3$, the eigenvalue of $\cC_6$ 
is expected to be given by a cubic polynomial in $\hbar^2g(g{-}1)$. 
We have evaluated this polynomial for the differential-operator realization, 
$p_i\mapsto\sfrac{\hbar}{\im}\partial_i$ and found it to be quadratic.

Finally, our methods can directly be used to find the two remaining expected Casimir operators for $W'_3$. 
It is only a matter of extending our ansatz below to sufficiently high polynomial order.
Unfortunately, we do not yet have arguments indicating their order or further restricting them.
Also, the methods presented here straightforwardly generalize to a larger number $N>3$ of particles,
except that multiple $sl(2)$ representations appear at fixed level four and higher~\cite{IsaLei}.
Again, we have no clue about the order of the corresponding Casimir operators. 
Otherwise, only the combinatorial complexity of the computations increases, 
making the use of computer algebra indispensable.


\section{Calogero invariants and their algebra}

\noindent
The quantum phase space of the rational $N$-particle Calogero model~\cite{Cal},
defined by the Hamiltonian
\begin{equation} \label{calham}
H\=\sfrac12\sum_i p_i^2\ +\ \sum_{i<j}\frac{\hbar^2g(g{-}1)}{(x_i{-}x_j)^2}\ ,
\end{equation}
is spanned by the particle coordinates~$x_i$ and their conjugate
momenta~$p_j$, with $i,j=1,2,\ldots,N$, and the canonical commutator
$[x_i, p_j]=\im\hbar\,\delta_{ij}$.
The particle masses have been scaled to unity, 
and the only parameter is the dimensionless Calogero parameter~$g\in\R$.
It is also useful to introduce the center-of-mass momentum and coordinate,
\begin{equation} \label{calmom}
P\=\sum_{i=1}^N p_i \und X\=\frac1N\sum_{i=1}^N x_i\ ,
\end{equation}
which form a subalgebra, $[X,P]=\im\hbar$, of any $W_N$ algebra.

The standard Liouville constants of motion for any value of~$g$
can be constructed from powers of the Dunkl operators~\cite{Dunkl,DunklXu}
\begin{equation} \label{dunkl}
\pi_i \= p_i + \im\sum_{j(\neq i)} \frac{\hbar\,g}{x_i{-}x_j}s_{ij} 
\end{equation}
where $s_{ij}=s_{ji}$ are the two-particle permutation operators, satisfying
$s_{ij}x_i=x_js_{ij}$,  $s_{ij}p_i=p_js_{ij}$ and  $s_{ij}^2=1$.
A set of algebraically independent Liouville integrals is given by
\begin{equation}
B_{0,k} \= \res\Bigl(\sum_i \pi_i^k\Bigr) \qquad\textrm{for}\quad k=1,2,\ldots,N\ ,
\end{equation}
with $\res(A)$ denoting the restriction of an operator~$A$ to the subspace of
states which are totally symmetric under any particle permutation.
Because the Dunkl operators commute, $[\pi_i,\pi_j]=0$,
it is easy to prove that the $B_{0,k}$ commute with one another,
\begin{equation} \label{Icom}
[B_{0,k},B_{0,\ell}]\=0\ .
\end{equation}
The first three Liouville integrals read
\begin{equation}
B_{0,1}\=P\ ,\qquad
B_{0,2}\=2H\ ,\qquad
B_{0,3}\=\sum_i p_i^3\ +\ 3\sum_{i<j}\frac{\hbar^2g(g{-}1)}{(x_i{-}x_j)^2}(p_i{+}p_j)\ ,
\end{equation}
where $P$ and $H$ are given in (\ref{calmom}) and (\ref{calham}), respectively.
Since (\ref{Icom}) contains in particular $[H,B_{0,k}]=0$,
one finds $N$ independent involutive constants of motion.
We also note the invariance under the involution $g\leftrightarrow 1{-}g$,
since the $g$-dependence comes in the combination~$g(g{-}1)$.

Together with
\begin{equation}
B_{1,1}\=\sfrac12\sum_i(x_ip_i+p_ix_i)\ =:\ D \und
B_{2,0}\=\sum_i x_i^2 \ =:\ 2K\ ,
\end{equation}
the Hamiltonian is part of an $sl(2)$ subalgebra,
\begin{equation}
\sfrac1\hbar[D,H]\=2\im H\ ,\qquad
\sfrac1\hbar[D,K]\=-2\im K\ ,\qquad
\sfrac1\hbar[K,H]\=\im D\ .
\end{equation}
This fact allows for the construction of many additional integrals,
from which one may choose $N{-}1$ functionally independent ones~\cite{Woj}.
One says that the Calogero model is maximally superintegrable.

Our goal is to investigate further the commutator algebra of 
permutation-invariant observables a.k.a.~one-particle operators
\begin{equation}
B_{k,\ell} \= \res\Bigl(\sum_i \weyl\bigl(x_i^k \pi_i^\ell\bigr)\Bigr)
\end{equation}
where ``weyl'' stands for Weyl ordering with respect to positions and momenta,
\begin{equation}
\weyl\bigl(x^k \pi^\ell\bigr) \= \sfrac{\pa^k}{\pa\alpha^k}\sfrac{\pa^\ell}{\pa\beta^\ell}\,\ep^{\alpha x+\beta\pi}\big|_{\alpha=\beta=0}
\qquad\Leftrightarrow\qquad
\ep^{\alpha x+\beta\pi} \= \sum_{k,\ell=0}^\infty 
\sfrac{\alpha^k}{k!}\sfrac{\beta^\ell}{\ell!}\,\weyl\bigl(x^k\pi^\ell\bigr)\ .
\end{equation}
We may organize this set of observables according to their total degree (in coordinates and Dunkl momenta)
denoted as the `level' $k{+}\ell$, and obtain an infinite pyramid
\begin{equation} \label{pyramid}
\begin{aligned}
B_{0,0}&=N \\[4pt]
B_{1,0}=NX \qquad & \qquad B_{0,1}=P \\[4pt]
B_{2,0}=2K \qquad\qquad B_{1,1}&=D \qquad\qquad B_{0,2}=2H \\[4pt]
B_{3,0} \qquad\qquad\qquad B_{2,1} \qquad\quad & \quad\qquad B_{1,2} \qquad\qquad\qquad B_{0,3} \\[4pt]
B_{4,0} \qquad\qquad\qquad B_{3,1} \qquad\qquad\qquad B&{}_{2,2} \qquad\qquad\qquad B_{1,3} \qquad\qquad\qquad B_{0,4} \\[4pt]
\cdots \qquad\qquad\qquad \cdots \qquad\qquad\qquad \cdots \qquad\quad & \quad\qquad \cdots \qquad\qquad\qquad \cdots \qquad\qquad\qquad \cdots
\end{aligned}
\end{equation}
whose $r$th row (starting from zero) contains the $r{+}1$ level-$r$ operators $B_{r,0},\ldots,B_{0,r}$.

Packaging these invariants into a generating function
\begin{equation}
B(\alpha,\beta) \= \sum_{k,\ell=0}^\infty \sfrac{\alpha^k}{k!}\sfrac{\beta^\ell}{\ell!}\,B_{k,\ell}
\= \res\Bigl(\sum_i \ep^{\alpha x_i+\beta\pi_i} \Bigr) \quad\for \alpha,\beta\in\R\ ,
\end{equation}
the computation of $[B(\alpha,\beta),B(\gamma,\delta)]$ requires the Baker--Campbell--Haussdorff formula
for the $S_N$-extended Heisenberg algebra generated by $x_i$ and~$\pi_j$,
\begin{equation}
\sfrac{1}{\im\hbar}\bigl[ x_i,\pi_j \bigr] \= \begin{cases}
1+\hbar g\sum_{k(\neq i)}s_{ik} & \for i=j \\[4pt]
-\hbar g s_{ij} & \for i\neq j \end{cases} \Biggr\} 
\qquad\qquad \textrm{and others commute}\ .
\end{equation}
It becomes manageable in the free-particle case $g{=}0$,
since for $\pi_i\to p_i$ one obtains
\begin{equation}
\sum_{i,j} \bigl[ \ep^{\alpha x_i+\beta p_i}\,,\, \ep^{\gamma x_j+\delta p_j} \bigr]
\= 2\im\,\sin\sfrac{\hbar}{2}(\alpha\delta{-}\beta\gamma) \sum_i \ep^{(\alpha+\gamma)x_i+(\beta+\delta)p_i} 
\end{equation}
and hence the `sine algebra'~\cite{FaZa}
\begin{equation}
\bigl[ B(\alpha,\beta)\,,\,B(\gamma,\delta)\bigr] \= 
2\im\,\sin\sfrac{\hbar}{2}(\alpha\delta{-}\beta\gamma)\ B(\alpha{+}\gamma,\beta{+}\delta)
\quad\for g{=}0\ ,
\end{equation}
from which we recover the $W_{1+\infty}$ algebra
\begin{equation} \label{Winfalgebra}
\sfrac{1}{\im\hbar}\,\bigl[ B_{k,\ell}\,,\,B_{m,n} \bigr] \= (kn{-}\ell m)\,B_{k+m-1,\ell+n-1}
\ +\ \sum_{r=1}^\infty \hbar^{2r} c^{2r+1}_{k\ell mn}\,B_{k+m-1-2r,\ell+n-1-2r}\ .
\end{equation}
The $c$-coefficients are read off by matching powers,
\begin{equation} \label{ccoeff}
c^{2r+1}_{k\ell mn} \= \sum_{s=0}^{2r+1} (-1)^{r+s} 
\frac{(k)_{2r+1-s}\,(\ell)_s\,(m)_s\,(n)_{2r+1-s}}{2^{2r} s!\,(2r{+}1{-}s)!}
\quad\with (x)_q=x(x{-}1)\cdots(x{-}q{+}1) \ ,
\end{equation}
which indeed yields $c^1_{k\ell mn}=kn{-}\ell m$ and
\begin{equation}
c^3_{k\ell mn} \= -\sfrac1{24}\bigl[ 
(k)_3(\ell)_0(m)_0(n)_3 - 3\,(k)_2(\ell)_1(m)_1(n)_2 + 3\,(k)_1(\ell)_2(m)_2(n)_1 - (k)_0(\ell)_3(m)_3(n)_0 \bigr]\ .
\end{equation}
This is all we shall need for the $W_3$ algebra at nonzero~$g$, as we explain shortly.

Three remarks are in order. Firstly, for widely-separated particle configurations, i.e.~$|x_i{-}x_j|\to\infty\ \forall i,j$,
the interaction contributions to $B_{k,\ell}$ go to zero, and hence the structure constants of our algebra cannot depend on the
Calogero parameter~$g$. The only exception is the constant central term~$B_{0,0}$, which in fact will be deformed by the
interaction.
More precisely, for $k{+}m=\ell{+}n=:2s{+}1$ an odd integer, 
the $r{=}s$ contribution in the sum on the right-hand side of (2.16) gets deformed,
\begin{equation} \label{gdef}
\hbar^{2s} c^{2s+1}_{k\ell mn}\,B_{0,0}
\qquad\Rightarrow\qquad
\hbar^{2s}\bigl( c^{2s+1}_{k\ell mn}\,B_{0,0} + P_r(g(g{-}1)) \bigr)
\end{equation}
where $P_s$ is a polynomial of order~$s$ (times the identity). 
Apart from this modification, the $W_{1+\infty}$ algebra \eqref{Winfalgebra} computed for the free case is also valid
for the Calogero system.
In other words, turning on the coupling~$g$ deforms the $W_{1+\infty}$ algebra~\eqref{Winfalgebra} with \eqref{ccoeff}
only as indicated in~\eqref{gdef}.
Secondly, rows one and two in~\eqref{pyramid} play a distinguished role. 
Their commutators with any $B_{k,\ell}$ are exactly given by
\begin{equation}
\begin{aligned}
\sfrac{1}{\im\hbar}\bigl[ B_{1,0},B_{m,n} \bigr] &= n\,B_{m,n-1} \ ,\quad
\sfrac{1}{\im\hbar}\bigl[ B_{0,1},B_{m,n} \bigr] = -m\,B_{m-1,n} \ , \\[4pt]
\sfrac{1}{\im\hbar}\bigl[ B_{2,0},B_{m,n} \bigr] = 2n\,B_{m+1,n-1} \ ,\quad &
\sfrac{1}{\im\hbar}\bigl[ B_{1,1},B_{m,n} \bigr] = (n{-}m)\,B_{m,n} \ ,\quad
\sfrac{1}{\im\hbar}\bigl[ B_{0,2},B_{m,n} \bigr] = -2m\,B_{m-1,n+1}\ .
\end{aligned}
\end{equation}
This shows that the level-one operators form the center-of-mass Heisenberg algebra, $[B_{1,0},B_{0,1}]=\im\hbar B_{0,0}$,
and that the level-2 operators are level preserving and span the above-mentioned $sl(2)$ subalgebra, 
in which $B_{1,1}{=}D$ is the grading operator (measuring the position of columns in~\eqref{pyramid}).
As a consequence, the $r$th row in~\eqref{pyramid} furnishes a spin-$\frac{r}2$ representation of $sl(2)$,
giving us an $sl(2)$ decomposition of the $W_{1+\infty}$ algebra.\footnote{
See also \cite{IsaLei,Hak}.}
Thirdly, if we fix the particle number~$N$ then all operators $B_{k,\ell}$ of level $k{+}\ell\,{>}\,N$ will be algebraically
dependent on those of level $k{+}\ell\le N$. 
For $N\ge3$ however, the commutators of lower-level operators produce such dependent operators in the algebra.
Expressing them in terms of polynomials of lower-level operators will turn the $W_{1+\infty}$ algebra 
into a nonlinear~$W_N$ algebra, which continues to respect the $sl(2)$ substructure. 
In this way, the $W_N$ algebra emerges from the Heisenberg algebra of the constituent particle coordinates and momenta.


\section{A $W_3$ algebra}

\noindent
As a warm-up, let us recall the two-particle case, $N{=}2$, which is still linear.
Its invariants are given by rows one and two in~\eqref{pyramid}. 
Their span $W_2$ is the above-mentioned $sl(2)$ conformal subalgebra together with a spin-$\half$ representation. 
Yet, inside the universal enveloping algebra the center-of-mass parts represented by the first row may be decoupled:
the combinations
\begin{equation} \label{B'2}
B'_{2,0} \= B_{2,0} - \half B_{1,0} B_{1,0}\ ,\quad
B'_{1,1} \= B_{1,1} - \sfrac14 \bigl( B_{1,0}B_{0,1}+B_{0,1}B_{1,0} \bigr)\ ,\quad
B'_{0,2} \= B_{0,2} - \half B_{0,1} B_{0,1}
\end{equation}
continue to form an $sl(2)$ algebra but commute with the center-of-mass parts $B_{1,0}$ and~$B_{0,1}$,
\begin{equation}
[B_{1,0},B'_{k,\ell}] \= 0 \= [B_{0,1},B'_{k,\ell}] \quad \for k{+}\ell=2\ ,
\end{equation}
allowing us to decompose $W_2=W'_2\oplus W_1=sl(2)\oplus\cH$, where $\cH$ denotes the Heisenberg algebra.
Such a decoupling will also be achieved for $N{=}3$ and brings a significant simplification, since 
the center-of-mass operators may be ignored henceforth. 
Nevertheless, there is one more invariant operator than phase-space degrees of freedom, hinting at the existence of a Casimir operator.
Clearly, the latter is given by the standard $sl(2)$ Casimir,
\begin{equation} \label{C2}
\cC_2 \= 2(K'H'+H'K')-D^{\prime\,2} \= \half( B'_{2,0}B'_{0,2}+B'_{0,2}B'_{2,0} ) - B_{1,1}^{\prime\,2}\ .
\end{equation}
Indeed, classically $\cC_2=0$ (free bosons or fermions) but quantum mechanically one may parametrize
\begin{equation}
\cC_2 \= \hbar^2 g(g{-}1) \quad\for g\in\R
\end{equation}
since unitarity demands that $\cC_2\ge-\sfrac14\hbar^2$.
This value is realized precisely by the Dunkl deformation~\eqref{dunkl}, 
which gives rise to the Calogero potential in the Hamiltonian~\eqref{calham}!
More generally, turning on the Calogero coupling in the free theory will deform
all invariant operators $B_{k,\ell}$ with $k\le\ell$ (though not $B_{1,1}$).

Let us now turn to the three-particle system, for which we have seven independent symmetric operators,
\begin{equation}
B_{1,0}\ ,\quad B_{0,1}\ ;\qquad B_{2,0}\ ,\quad B_{1,1}\ ,\quad B_{0,2}\ ;\qquad
B_{3,0}\ ,\quad B_{2,1}\ ,\quad B_{1,2}\ ,\quad B_{0,3}\ .
\end{equation}
The $r{=}3$ invariants take the form~\footnote{
At the cubic level, all symmetric orderings agree. This no longer holds for quartic invariants.}
\begin{equation}
\begin{aligned}
& B_{3,0}\=\sum_i x_i^3\ ,\quad
B_{2,1}\=\sum_i \weyl\bigl(x_i^2 p_i\bigr) 
\= \sfrac12 \sum_i \bigl(x_i^2 p_i + p_i\,x_i^2\bigr) \= \sum_i x_i p_i x_i\ , \\[4pt]
& B_{1,2}\=\sum_i \weyl\bigl(x_i\,p_i^2\bigr) + \sum_{i<j} \frac{\hbar^2 g(g{-}1)}{(x_i{-}x_j)^2}(x_i{+}x_j)\ ,\quad
B_{0,3}\=\sum_i p_i^3\ +\ 3\sum_{i<j}\frac{\hbar^2g(g{-}1)}{(x_i{-}x_j)^2}(p_i{+}p_j)\ .
\end{aligned}
\end{equation}
In search of a Casimir operator we will deal with the universal enveloping algebra ${\cal U}(W_3)$ 
and thus have to settle on an ordering convention for the generators $B_{k,\ell}$.
The most natural choice is Weyl ordering,
\begin{equation}
\Weyl\bigl(A_1 A_2 \cdots A_q\bigr) \ :=\ 
\sfrac{\pa}{\pa\alpha_1}\cdots\sfrac{\pa}{\pa\alpha_q}\,\ep^{\alpha_1 A_1 + \ldots + \alpha_q A_q}\big|_{\vec{\alpha}=0}
\= \sfrac{1}{q!}\!\!\sum_{\si\in S_q} A_{\si(1)} A_{\si(2)} \cdots A_{\si(q)} \quad\for A_s \in \{ B_{k,\ell} \}\ ,
\end{equation}
where the sum runs over all permutations~$\si$ of the symmetric group~$S_q$.
It is important to note that this Weyl ordering does {\it not\/} refer to positions and momenta,
but only to the operators $B_{k,\ell}$ themselves, so it differs from the one introduced before!
To streamline notation, we abbreviate~\footnote{
To illustrate this Weyl ordering, 
$(21|10)=\tfrac12(B_{2,1}B_{1,0}+B_{1,0}B_{2,1})$, for example.}
\begin{equation}
B_{k,\ell}\ =:\ (k\ell) \und \Weyl(B_{k,\ell}B_{m,n}\ldots B_{s,t})\ =:\ (k\ell|mn|\ldots|st)\ .
\end{equation}

The $N{=}3$ system adds the row $r{=}3$ to the previous algebra, which transforms in a spin-$\sfrac32$
representation of the conformal $sl(2)$. The only nontrivial commutators are those
of level-3 operators with each other, which we display in Table~\ref{tab1}.
\begin{table}[h!]
\begin{center}
\caption{}
\label{tab1}
\begin{tabular}{|c||c|c|c|c|}
\hline
$\vphantom{\Big|}\sfrac{1}{\im\hbar}[B_{k,\ell},B_{m,n}]$ & (30) & (21) & (12) & (03) \\
\hline\hline$\vphantom{\Big|}$
\multirow{2}{*}{(30)} & \multirow{2}{*}{0} & \multirow{2}{*}{3(40)} & 
\multirow{2}{*}{6(31)} & $9(22)-\sfrac32\hbar^2(00)$ \\
$\vphantom{\Big|}$ & & & & $+9\hbar^2g(g{-}1)$ \\
\hline$\vphantom{\Big|}$
\multirow{2}{*}{(21)} & \multirow{2}{*}{$-3(40)$} & \multirow{2}{*}{0} & 
$3(22)+\sfrac12\hbar^2(00)$ & \multirow{2}{*}{6(13)} \\
$\vphantom{\Big|}$ & & & $-3\hbar^2g(g{-}1)$ & \\
\hline$\vphantom{\Big|}$
\multirow{2}{*}{(12)} & $-6(31)$ & $-3(22)-\sfrac12\hbar^2(00)$ & 
\multirow{2}{*}{0} & \multirow{2}{*}{3(04)} \\
$\vphantom{\Big|}$ & & $+3\hbar^2g(g{-}1)$ & & \\
\hline$\vphantom{\Big|}$
\multirow{2}{*}{(03)} & $-9(22)+\sfrac32\hbar^2(00)$ & 
\multirow{2}{*}{$-6(13)$} & \multirow{2}{*}{$-3(04)$} & \multirow{2}{*}{0} \\
$\vphantom{\Big|}$ & $-9\hbar^2g(g{-}1)$ & & & \\
\hline
\end{tabular}
\end{center}
\end{table}
We see that the Calogero coupling explicitly appears only in the central terms.
The level-4 operators are expressed in terms of lower-level invariants,
\begin{equation} \label{level4}
\begin{aligned}
(40) &\= \sfrac43(30|10)+\sfrac12(20|20)-(20|10|10)+\sfrac16(10|10|10|10)\ ,\\[4pt]
(31) &\= \sfrac13(30|01)+(21|10)+\sfrac12(20|11)-\sfrac12(20|10|01)-\sfrac12(11|10|10)+\sfrac16(10|10|10|01)\ , \\[4pt]
(22) &\= \sfrac23(21|01)+\sfrac16(20|02)+\sfrac23(12|10)+\sfrac13(11|11)-\sfrac16(20|01|01) \\
& \qquad\qquad\qquad\qquad\qquad -\sfrac23(11|10|01)-\sfrac16(10|10|02)+\sfrac16(10|10|01|01) \ , \\[4pt]
(13) &\= (12|01)+\sfrac12(11|02)+\sfrac13(10|03)-\sfrac12(11|01|01)-\sfrac12(10|02|01)+\sfrac16(10|01|01|01)\ , \\[4pt]
(04) &\= \sfrac43(03|01)+\sfrac12(02|02)-(02|01|01)+\sfrac16(01|01|01|01)\ .
\end{aligned}
\end{equation}
With this input, the above table becomes the nonlinear commutation relations of a $W_3$ algebra.

Once we have restricted ourselves to the set $\{(10),(01);(20),(11),(02);(30),(21),(12),(03)\}$,
it is straightforward to construct unique nonlinear combinations which commute with $\{(10),(01)\}$:
\begin{equation} \label{Bprime}
\begin{aligned}
B'_{2,0}\ \equiv\ (20)' &\= (20) -\sfrac13(10|10)\ ,\\
B'_{1,1}\ \equiv\ (11)' &\= (11) -\sfrac13(10|01)\ ,\\
B'_{0,2}\ \equiv\ (02)' &\= (02) -\sfrac13(01|01)\ ,\\
B'_{3,0}\ \equiv\ (30)' &\= (30) -(20|10) +\sfrac29(10|10|10)\ ,\\
B'_{2,1}\ \equiv\ (21)' &\= (21) -\sfrac13(20|01) -\sfrac23(11|10) +\sfrac29(10|10|01)\ ,\\
B'_{1,2}\ \equiv\ (12)' &\= (12) -\sfrac23(11|01) -\sfrac13(10|02) +\sfrac29(10|01|01)\ ,\\
B'_{0,3}\ \equiv\ (03)' &\= (03) -(02|01) +\sfrac29(01|01|01)\ .
\end{aligned}
\end{equation}
The decoupled operators $\{(k\ell)'\}$ with $k{+}\ell=2$ or~$3$ span the smaller algebra $W'_3$.
Its commutators involving level-2 operators remain unchanged, but the nontrivial level-3 commutators
get modified. This is a nontrivial computation, because resolving $(k\ell)$ in terms of $(k\ell)'$ via \eqref{Bprime}
and inserting this into \eqref{level4} yields Weyl-ordered products as factors inside another Weyl-ordered product,
which then has to be re-expressed in terms of Weyl-ordered products of the original operators
and their iterated commutators. In particular, one has to employ
\begin{equation} \label{reorder}
\begin{aligned}
\bigl(a\big|(b|c)\bigr) &\= (a|b|c) + \sfrac1{12} \bigl\{ [[a,b],c]+[[a,c],b] \bigr\}\ ,\\[4pt]
\bigl(a\big|b\big|(c|d)\bigr) &\= (a|b|c|d) +\sfrac1{12} \bigl\{
\bigl(a\big|[[b,c],d]\bigr)+\bigl(a\big|[[b,d],c]\bigr)+
\bigl([a,c]\big|[b,d]\bigr)+ \ (a\leftrightarrow b) \bigr\}\ ,\\[4pt]
\bigl(a\big|(b|c|d)\bigr) &\= (a|b|c|d) + \sfrac{1}{12}\bigl\{
\bigl(b\big|[[a,c],d]\bigr)+\bigl(b\big|[[a,d],c]\bigr)+\ \textrm{cyclic in $(b,c,d)$}\bigr\}\ ,\\[4pt]
\bigl((a|b)\big|(c|d)\bigr) &\= (a|b|c|d) + \sfrac{1}{12}\bigl\{
\bigl(a\big|[[b,c],d]\bigr)+\bigl(a\big|[[b,d],c]\bigr)+\bigl(b\big|[[a,c],d]\bigr)+\bigl(b\big|[[a,d],c]\bigr)
+\bigl({\textstyle{a\leftrightarrow c \atop b\leftrightarrow d}}\bigr)\bigr\}\\
&\qquad\qquad\quad\ \ +\sfrac{1}{4}\bigl\{\bigl([a,c]\big|[b,d]\bigr)+\bigl([a,d]\big|[b,c]\bigr)\bigr\}\ ,
\end{aligned}
\end{equation}
where $(a\,{\leftrightarrow}\,b)$ means adding a copy of all previous terms with $a$ and $b$ interchanged,
and ``$\textrm{cyclic in $(b,c,d)$}$'' instructs to add to each previous term two more obtained by cyclicly permuting
the labels $b$, $c$ and~$d$.
One sees that recasting the iterated Weyl ordering to simple Weyl ordering always produces two commutators,
i.e.~quantum corrections of order~$\hbar^2$.
After putting $(00)=3$, the result is displayed in Table~\ref{tab2}, 
where we use the short-hand $(k\ell|mn|\ldots|st)':=\Weyl(B'_{k,\ell}B'_{m,n}\ldots B'_{s,t})$.
\begin{table}[h!]
\begin{center}
{\small
\caption{}
\label{tab2}
\begin{tabular}{|c||c|c|c|c|}
\hline
$\vphantom{\Big|}\!\!\sfrac{1}{\im\hbar}[B'_{k,\ell},B'_{m,n}]\!\!$ & $(30)'$ & $(21)'$ & $(12)'$ & $(03)'$ \\
\hline\hline$\vphantom{\Big|}$
\multirow{2}{*}{$(30)'$} & \multirow{2}{*}{0} & \multirow{2}{*}{$\sfrac12(20|20)'$} & 
\multirow{2}{*}{$(20|11)'$} & $\!\!-\sfrac32(20|02)'+3(11|11)'\!\!$ \\
$\vphantom{\Big|}$ & & & & $+\hbar^2[9g(g{-}1)-4]$ \\
\hline$\vphantom{\Big|}$
\multirow{2}{*}{$(21)'$} & \multirow{2}{*}{$-\sfrac12(20|20)'$} & \multirow{2}{*}{0} & 
$\!\!\sfrac56(20|02)'-\sfrac13(11|11)'\!\!$ & \multirow{2}{*}{$(11|02)'$} \\
$\vphantom{\Big|}$ & & & $-\hbar^2[3g(g{-}1)-\sfrac43]$ & \\
\hline$\vphantom{\Big|}$
\multirow{2}{*}{$(12)'$} & \multirow{2}{*}{$-(20|11)'$} & $\!\!-\sfrac56(20|02)'+\sfrac13(11|11)'\!\!$ 
& \multirow{2}{*}{0} & \multirow{2}{*}{$\sfrac12(02|02)'$} \\
$\vphantom{\Big|}$ & & $+\hbar^2[3g(g{-}1)-\sfrac43]$ & & \\
\hline$\vphantom{\Big|}$
\multirow{2}{*}{$(03)'$} & $\!\!\sfrac32(20|02)'-3(11|11)'\!\!$ & \multirow{2}{*}{$-(11|02)'$} & 
\multirow{2}{*}{$-\sfrac12(02|02)'$} & \multirow{2}{*}{0} \\
$\vphantom{\Big|}$ & $-\hbar^2[9g(g{-}1)-4]$ & & & \\
\hline
\end{tabular}
}
\end{center}
\end{table}

It is revealing to rewrite the $W'_3$ algebra in $sl(2)$ covariant notation.
To this end, we relabel the generators such as to bring out their $sl(2)$ transformation properties,
\begin{equation}
\begin{aligned}
&(20)' \= \sqrt{8}\,J_{-1}\ ,\qquad 
(11)' \= 2\,J_0\ ,\qquad
(02)' \= \sqrt{8}\,J_{+1}\ ,\\[4pt]
(30)' \= 2\,K_{-3/2}&\ ,\qquad
(21)' \= \sfrac{2}{\sqrt{3}}\,K_{-1/2}\ ,\qquad
(12)' \= \sfrac{2}{\sqrt{3}}\,K_{+1/2}\ ,\qquad
(03)' \= 2\,K_{+3/2}\ .
\end{aligned}
\end{equation}
The spin-$1$ representation $\{J_i\}$ with $i\in\{-1,0,1\}$ and the spin-$\sfrac32$ one
$\{K_\alpha\}$ with $\alpha\in\{-\sfrac32,-\sfrac12.\sfrac12,\sfrac32\}$ feature standard
descent chains,
\begin{equation}
\sfrac1{\im\hbar}\textrm{ad}\,J_{-1}: \quad J_{+1}\ \mapsto\ J_0\ \mapsto\ J_{-1}\ \mapsto\ 0 \und
K_{+3/2}\ \mapsto\ \sqrt{\vphantom{|}\smash{\sfrac32}}\,K_{+1/2}\ \mapsto\ 
\sqrt{3}\,K_{1/2}\ \mapsto\ \sfrac{3}{\sqrt{2}}\,K_{-3/2}\ \mapsto\ 0\ ,
\end{equation}
and $J_0$ has been normalized to obtain its conventional eigenvalues,
\begin{equation}
\sfrac1{\im\hbar}\textrm{ad}\,J_0: \quad J_i\ \mapsto\ i\,J_i \und K_\alpha\ \mapsto\ \alpha\,K_\alpha\ .
\end{equation}
The free system ($g(g{-}1){=}0$) also features a symplectic conjugation,
\begin{equation}
x_i^\+=p_i\ ,\quad p_i^\+=-x_i \quad\Rightarrow\quad
B_{k,\ell}^\+ = (-1)^\ell B_{\ell,k}\ ,
\end{equation}
thus 
\begin{equation}
J_0^\+=-J_0\ ,\quad J_{\pm1}^\+ = J_{\mp1} \und
K_{\pm 3/2}^\+ = \mp K_{\mp 3/2}\ ,\quad K_{\pm 1/2}^\+ = \pm K_{\mp 1/2}\ ,
\end{equation}
which is broken by the interaction.

The part of the $W'_3$ algebra involving the spin-1 generators is standard,
\begin{equation}
\sfrac{1}{\im\hbar}[J_i,J_k] \= f_{ik}^{\ \ell}\,J_\ell \und 
\sfrac{1}{\im\hbar}[J_i,K_\alpha] \= f_{i\alpha}^{\ \beta}\,K_\beta\ ,
\end{equation}
where $\{f_{ik}^{\ \ell}\}$ are the $sl(2)$ structure constants and
$\{f_{i\alpha}^{\ \beta}\}$ give the components of the spin-$\sfrac32$ representation matrices~$f_i$.
The $[K,K]$ commutators are more interesting. 
Their left-hand side transforms in an antisymmetric tensor-square of the spin-$\sfrac32$
representation, while the right-hand side sits in a symmetric tensor-square of the adjoint
(spin-$1$) representation. Both sides match,
\begin{equation}
\bigl[ {\bf\sfrac32} \otimes {\bf \sfrac32} \bigr]_{\textrm{A}} \=
{\bf 2} \oplus {\bf 0} \= \bigl[ {\bf 1} \otimes {\bf 1} \bigr]_{\textrm{S}}\ ,
\end{equation}
and hence the $[K,K]$ commutator provides a spin-2 representation $\{L_A\}$
with $A\in\{-2,-1,0,1,2\}$ and a singlet, which must be a linear combination of the
$sl(2)$ Casimir $\cC_2$ and a central term $\in\R$. The descent chain for the
spin-2 representation reads
\begin{equation}
\sfrac1{\im\hbar}\textrm{ad}\,J_{-1}: \quad (J_{+1}|J_{+1})\ \mapsto\ 2(J_{+1}|J_0)\ \mapsto\  
2\bigl\{(J_{+1}|J_{-1})+(J_0|J_0)\bigr\}\ \mapsto\ 6(J_0|J_{-1})\ \mapsto\ 6(J_{-1}|J_{-1})
\ \mapsto\ 0\ ,
\end{equation}
which is orthogonal to the Casimir singlet 
\begin{equation}
\cC_2 \= (20|02)'-(11|11)' \= 8\bigl(J_{+1}\big|J_{-1}\bigr)-4\bigl(J_0\big|J_0\bigr)\ .
\end{equation}
Altogether, Table~\ref{tab2} is then reassembled in Table~\ref{tab3},
which nicely demonstrates that the quantum corrections are $sl(2)$ invariant and
spanned by $\cC_2$ and a $g$-dependent central part.
\begin{table}[h!]
\begin{center}
{\small
\caption{}
\label{tab3}
\begin{tabular}{|c||c|c|c|c|}
\hline
$\vphantom{\Big|}\!\!\sfrac{1}{\im\hbar}[K_\alpha,K_\beta]\!\!$ & $K_{3/2}$ & $K_{1/2}$ & $K_{-1/2}$ & $K_{-3/2}$ \\
\hline\hline$\vphantom{\Big|}$
\multirow{2}{*}{$K_{3/2}$} & \multirow{2}{*}{0} & \multirow{2}{*}{$-\sqrt{3}\,(J_{+1}|J_{+1})$} &
\multirow{2}{*}{$-\sqrt{6}\,(J_{+1}|J_0)$} & $-(J_{+1}|J_{-1})-(J_0|J_0)$ \\
$\vphantom{\Big|}$ & & & & $\!\!-\sfrac12\cC_2-\hbar^2[\sfrac94g(g{-}1){-}1]\!\!$ \\
\hline$\vphantom{\Big|}$
\multirow{2}{*}{$K_{1/2}$} & \multirow{2}{*}{$\sqrt{3}\,(J_{+1}|J_{+1})$} & \multirow{2}{*}{0} &
$-(J_{+1}|J_{-1})-(J_0|J_0)$ & \multirow{2}{*}{$-\sqrt{6}\,(J_0|J_{-1})$} \\
$\vphantom{\Big|}$ & & & $\!\!+\sfrac12\cC_2+\hbar^2[\sfrac94g(g{-}1){-}1]\!\!$ & \\
\hline$\vphantom{\Big|}$
\multirow{2}{*}{$K_{-1/2}$} & \multirow{2}{*}{$\sqrt{6}\,(J_{+1}|J_0)$} & $(J_{+1}|J_{-1})+(J_0|J_0)$ &
\multirow{2}{*}{0} & \multirow{2}{*}{$-\sqrt{3}\,(J_{-1}|J_{-1})$} \\
$\vphantom{\Big|}$ & & $\!\!-\sfrac12\cC_2-\hbar^2[\sfrac94g(g{-}1){-}1]\!\!$ & & \\
\hline$\vphantom{\Big|}$
\multirow{2}{*}{$K_{-3/2}$} & $(J_{+1}|J_{-1})+(J_0|J_0)$ & \multirow{2}{*}{$\sqrt{6}\,(J_0|J_{-1})$} &
\multirow{2}{*}{$\sqrt{3}\,(J_{-1}|J_{-1})$} & \multirow{2}{*}{0} \\
$\vphantom{\Big|}$ & $\!\!+\sfrac12\cC_2+\hbar^2[\sfrac94g(g{-}1){-}1]\!\!$ & & & \\
\hline
\end{tabular}
}
\end{center}
\end{table}
In a nutshell, the commutator reads
\begin{equation} \label{W3cov}
\sfrac{1}{\im\hbar}[K_\alpha,K_\beta] 
\= f_{\alpha\beta}^{\ A}\,L_A + \epsilon_{\alpha\beta}\bigl(\sfrac12\,\cC_2+\hbar^2[\sfrac94g(g{-}1){-}1]\bigr)
\= f_{\alpha\beta}^{i\,k}\,(J_i|J_k) + \epsilon_{\alpha\beta}\bigl(\sfrac12\,\cC_2+\hbar^2[\sfrac94g(g{-}1){-}1]\bigr)\ ,
\end{equation}
where $f_{\alpha\beta}^{\ A}$ are the coupling coefficients for ${\bf\sfrac32}\otimes{\bf\sfrac32}\mapsto{\bf 2}$,
and $f_{\alpha\beta}^{i\,k}=f_{\alpha\beta}^A f_A^{i\,k}$ couples 
$\bigl[ {\bf\sfrac32} \otimes {\bf \sfrac32} \bigr]_{\textrm{A}}$ directly to the traceless quadratic form 
$\bigl[ {\bf 1} \otimes {\bf 1} - \textrm{trace}\bigr]_{\textrm{S}}$.
The singlet ${\bf\sfrac32}\otimes{\bf\sfrac32}\mapsto{\bf 0}$ provides an antisymmetric metric
$\epsilon_{\alpha\beta}=\pm\delta_{\alpha+\beta,0}$
in the spin-$\sfrac32$ module.

How rigid is this algebra? 
One is of course free to an overall rescaling of $\{J_i\}$ and also of $\{K_\alpha\}$, which we have employed to
normalize $\{f_{ik}^{\ \ell}\}$ and $\{f_{\alpha\beta}^{i\,k}\}$. The matrices $(f_i)_\alpha^\beta$
are then fixed by representation theory, as well as the antisymmetric ``metric''~$\epsilon_{\alpha\beta}$.
Hence, the $sl(2)$ decomposition of our algebra determines almost all structure constants of our algebra.
The only a priori undetermined part is the singlet piece, i.e.~the coefficients between the round brackets
in~\eqref{W3cov}. Incidentally, both the quantum ($\hbar$) and the Calogero ($g$) deformation from the
free classical Poisson algebra, $\sfrac1{\im\hbar}[.,.]\to\{.,.\}$, appear only in this place.


\section{A Casimir operator}

\noindent
The key task of this work is the construction of a Casimir operator for the nonlinear algebra $W'_3$ and thus also for~$W_3$.
It is clear that $\cC_2$ cannot do the job since it does not commute with the level-3 operators. 
Trial and error shows that the smallest candidate for such a Casimir operator must be a polynomial in $B'_{k,\ell}$ 
of degree at least six, with components transforming in representations of $sl(2)$-spin up to six or more.
Ignoring any quantum corrections, i.e.~setting $\hbar=0$ for the moment, the most general ansatz
for such a Casimir operator of (minimal) level 12 reads
\begin{equation} \label{ansatzclass}
\cC_6^{\textrm{class}} \ :=\ \alpha\,T^{\prime 6}_{66} + \beta\,T^{\prime 5}_{66} + \gamma\,T^{\prime 4}_{66} \qquad\with \alpha,\beta,\gamma\in\R\ ,
\end{equation}
where $T^{\prime 6}_{66}$ is a linear combination of Weyl-symmetrized products of six level-2 operators, 
$T^{\prime 5}_{66}$ is a linear combination of Weyl-symmetrized products of three level-2 and two level-3 operators,
and $T^{\prime 4}_{66}$ is  a linear combination of Weyl-symmetrized products of four level-3 operators,
in such a way that their first index ($k$) adds up to 6, as well as their second index ($\ell$).
This equality ensures that the ansatz commutes with~$(11)'$.
One may count that $T^{\prime 6}_{66}$ is composed of 4 terms, $T^{\prime 5}_{66}$ is made of 16 terms, and $T^{\prime 4}_{66}$ comprises 5 terms.
Because the adjoint action of level-2 operators is level-preserving, commutators with $(20)'$ or $(02)'$
do not relate the three pieces in the ansatz~\eqref{ansatzclass}. Therefore, the requirements
\begin{equation}
\bigl[(20)',T^{\prime s}_{66}\bigr] \= \bigl[(02)',T^{\prime s}_{66}\bigr] \= 0 \quad\for s=6,5,4
\end{equation}
can be used to fix the relative coefficients of the various contributions inside each~$T^{\prime s}_{66}$.
The solution of this linear problem is unique (up to scale) for $s{=}6$ and $s{=}4$, but leaves 
a one-parameter family (plus overall scale) for $s{=}5$. Further conditions from vanishing commutators 
with level-3 operators will remove this ambiguity. Anticipating this result, as an intermediate
concretization we obtain 
\begin{equation} \label{T66}
\begin{aligned}
T^{\prime 6}_{66} &\= (20|20|20|02|02|02)'-3\,(20|20|11|11|02|02)'+3\,(20|11|11|11|11|02)'-(11|11|11|11|11|11)'\ ,\\[4pt]
T^{\prime 5}_{66} &\= (30|30|02|02|02)'-6\,(30|21|11|02|02)'+6\,(30|20|12|02|02)'-6\,(30|20|11|03|02)' \\
&\ +4\,(30|11|11|11|03)'-3\,(21|21|20|02|02)'+12\,(21|21|11|11|02)'+6\,(21|20|20|03|02)'\\
&\ -6\,(21|20|12|11|02)'-12\,(21|12|11|11|11)'+(20|20|20|03|03)'-3\,(20|20|12|12|02)'\\
&\ -6\,(20|20|12|11|03)'+12\,(20|12|12|11|11)'\ ,\\[4pt]
T^{\prime 4}_{66} &\= (30|30|03|03)'-6\,(30|21|12|03)'+4\,(30|12|12|12)'+4\,(21|21|21|03)'-3\,(21|21|12|12)' \ ,
\end{aligned}
\end{equation}
where the overall scale of each $T^{\prime s}_{66}$ was fixed such that the first term has weight~1.
Of the 16 possible terms for $T^{\prime 5}_{66}$ only 14 appear, because the vanishing condition for the
commutators with level-3 operators will enforce zero weight for $(30|12|11|11|02)'$ and $(21|20|11|11|03)'$.

Since adding a single level-3 operator suffices to generate the full $W'_3$ algebra from $sl(2)$,
we only need to impose the vanishing of $[\cC_6^{\textrm{class}},(30)']$.
For the classical result, the ordering inside operator products is irrelevant, and so we may ignore
the reordering issues displayed in~\eqref{reorder}. Employing the classical part of Table~\ref{tab2},
this commutator generates the following linear combinations (indicated by `$\&$'),
\begin{equation}
\sfrac1{\im\hbar}\bigl[ T^{\prime 6}_{66}, (30)' \bigr] \ \buildrel{\hbar=0}\over\longrightarrow\ T^{\prime 6}_{85}\ ,\qquad
\sfrac1{\im\hbar}\bigl[ T^{\prime 5}_{66}, (30)' \bigr] \ \buildrel{\hbar=0}\over\longrightarrow\ T^{\prime 5}_{85} \ \&\ T^{\prime 6}_{85}\ ,\qquad
\sfrac1{\im\hbar}\bigl[ T^{\prime 4}_{66}, (30)' \bigr] \ \buildrel{\hbar=0}\over\longrightarrow\ T^{\prime 5}_{85}\ ,
\end{equation}
and no terms of the form $T^{\prime 4}_{85}$. Cancellation of all resulting structures on the
right-hand side is an overdetermined linear problem, which however has a unique (up to scale) solution,
\begin{equation}
(\alpha,\beta,\gamma) \= (6,9,-54)\ .
\end{equation}
The expression $\cC_6^{\textrm{class}}$ in \eqref{ansatzclass} hence Poisson-commutes with all
generators of~$W'_3$. We note that this result does not see the $g$~dependence of Table~\ref{tab2}
or Table~\ref{tab3}.

Let us then turn on $\hbar$ and take into account the quantum corrections. 
This is much more complicated, mostly due to the Weyl reordering \`a la~\eqref{reorder}. 
However now we require formul\ae\ for $\big(a\big|b\big|c\big|(d|e)\bigr)$ and
$\big(a\big|b\big|c\big|d\big|(e|f)\bigr)$, which we refrain from displaying here.
Furthermore, the quantum corrections to the commutators $\sfrac1{\im\hbar}\bigr[T^{\prime s}_{66},(30)'\bigl]$ 
are $g$-dependent and of lower degree in the $(k\ell)'$,
\begin{equation} \label{30com66}
\sfrac1{\im\hbar}\bigl[T^{\prime 5}_{66},(30)'\bigr]\ \longrightarrow\ 
T^{\prime 5}_{85}\ \&\ T^{\prime 6}_{85}\ \&\ \hbar^2 T^{\prime 4}_{63}\ \&\ \hbar^4 T^{\prime 2}_{41}\ ,\qquad
\sfrac1{\im\hbar}\bigl[T^{\prime 4}_{66},(30)'\bigr]\ \longrightarrow\ 
T^{\prime 5}_{85}\ \&\ \hbar^2 T^{\prime 3}_{63}\ \&\ \hbar^4 T^{\prime 2}_{41}\ .
\end{equation}
Clearly, already the $O(\hbar^2)$ contributions cannot cancel between $T^{\prime 5}_{66}$ and $T^{\prime 6}_{66}$,
making a quantum deformation of the classical Casimir operator~\eqref{ansatzclass} unavoidable,
\begin{equation} \label{ansatzquant}
\cC_6^{\textrm{quant}} \ :=\ \alpha\,T^{\prime 6}_{66} + \beta\,T^{\prime 5}_{66} + \gamma\,T^{\prime 4}_{66}
+ \delta\,\hbar^2 T^{\prime 4}_{44} + \epsilon\,\hbar^2 T^{\prime 3}_{44} + \zeta\,\hbar^4 T^{\prime 2}_{22}\ ,
\end{equation}
in obvious notation and with three more coefficients $(\delta,\epsilon,\zeta)$ to be determined.
More explicitly,
\begin{equation} \label{T44}
\begin{aligned}
T^{\prime 4}_{44} &\= (20|20|02|02)'-2\,(20|11|11|02)'+(11|11|11|11)'\ ,\\[4pt]
T^{\prime 3}_{44} &\= (30|12|02)'-(30|11|03)'-(21|21|02)'+(21|20|03)'+(21|12|11)'-(20|12|12)'\ ,\\[4pt]
T^{\prime 2}_{22} &\= (20|02)'-(11|11)'\ .
\end{aligned}
\end{equation}
Meticulous computations of the commutators $\sfrac1{\im\hbar}\bigr[T^{\prime s}_{44},(30)'\bigl]$ and
$\sfrac1{\im\hbar}\bigr[T^{\prime 2}_{22},(30)'\bigl]$ produce
\begin{equation}
\sfrac1{\im\hbar}\bigl[ T^{\prime 4}_{44}, (30)' \bigr] \ \longrightarrow\ T^{\prime 4}_{63}\ ,\qquad
\sfrac1{\im\hbar}\bigl[ T^{\prime 3}_{44}, (30)' \bigr] \ \longrightarrow\ T^{\prime 4}_{63} \ \&\ T^{\prime 3}_{63}\ \&\ \hbar^2 T^{\prime 2}_{41}\ ,\qquad
\sfrac1{\im\hbar}\bigl[ T^{\prime 2}_{22}, (30)' \bigr] \ \longrightarrow\ T^{\prime 2}_{41}\ ,
\end{equation}
which can be matched in structure with~\eqref{30com66}. Still, the system of vanishing commutator equations 
is overdetermined, but miraculously it admits the solution
\begin{equation}
(\alpha,\beta,\gamma,\delta,\epsilon,\zeta) \= 
\bigl(6\,,\,9\,,\,-54\,,\,207{-}108g(g{-}1)\,,\,648{-}324g(g{-}1)\,,\,709{-}1656g(g{-}1){+}486g^2\!(g{-}1)^2\bigr)\ .
\end{equation}
With these coefficient values, the degree-6 polynomial (in $B'_{k,\ell}$) given in~\eqref{ansatzquant}
together with \eqref{T66} and~\eqref{T44} commutes with all generators of~$W'_3$ 
and is thus a first Casimir operator of that algebra. 
Combining the previous formul\ae\ and abbreviating $g(g{-}1)=\lambda$, in a single expression it reads 
\begin{equation}
\begin{aligned} \label{Cquant}
\cC_6^{\textrm{quant}} 
&\,\= 6\,\bigl\{ (20|20|20|02|02|02)'-3\,(20|20|11|11|02|02)'+3\,(20|11|11|11|11|02)'-(11|11|11|11|11|11)' \bigr\} \\[2pt]
&\quad +9\,\bigl\{ (30|30|02|02|02)'-6\,(30|21|11|02|02)'+6\,(30|20|12|02|02)'-6\,(30|20|11|03|02)' \\
&\qquad +4\,(30|11|11|11|03)'-3\,(21|21|20|02|02)'+12\,(21|21|11|11|02)'+6\,(21|20|20|03|02)'\\
&\qquad -6\,(21|20|12|11|02)'-12\,(21|12|11|11|11)'+(20|20|20|03|03)'-3\,(20|20|12|12|02)'\\
&\qquad -6\,(20|20|12|11|03)'+12\,(20|12|12|11|11)' \bigr\} \\[2pt]
&\quad -54\,\bigl\{(30|30|03|03)'-6\,(30|21|12|03)'+4\,(30|12|12|12)'+4\,(21|21|21|03)'-3\,(21|21|12|12)'\bigr\} \\[2pt]
&\quad +9(23{-}12\lambda)\hbar^2 \bigl\{ (20|20|02|02)'-2\,(20|11|11|02)'+(11|11|11|11)' \bigr\} \\[2pt]
&\quad +324(2{-}\lambda)\hbar^2 \bigl\{ (30|12|02)'-(30|11|03)'-(21|21|02)'+(21|20|03)'+(21|12|11)'-(20|12|12)' \bigr\} \\[2pt]
&\quad +(709{-}1656\lambda{+}486\lambda^2)\hbar^4 \bigl\{ (20|02)'-(11|11)' \bigr\}\ .
\end{aligned}
\end{equation}

In addition, it provides a Casimir operator for the bigger algebra~$W_3$ (including the center-of-mass
operators) upon inserting~\eqref{Bprime} and again Weyl reordering.
This form was obtained using {\tt Mathematica} and is displayed in the Appendix below.
On the Calogero wave functions, the operators~$B_{k,\ell}$ are realized by partial differential operators.
In this differential-operator realization of the $W_3$ algebra, putting $p_i\mapsto \sfrac{\hbar}{\im}\partial_i$,
our Casimir operator is proportional to the identity, taking the following value,
\begin{equation}
\cC_6^{\textrm{quant}}\ \mapsto\ ( 144 + 216\,\lambda - 1215\,\lambda^2 )\,\hbar^6
\qquad\with\quad \lambda=g(g{-}1)\ .
\end{equation}


\appendix
\section{Appendix}
Putting back the unprimed operators (\ref{Bprime}) into (\ref{Cquant}) and Weyl-reordering we get
\begin{multline} 
{\cC}_6^{\textrm{quant}}\=
	3T^9_{66}-3T^8_{66}+9T^7_{66}-3T^6_{66}+9T^5_{66}-54T^4_{66}\\[4pt]
	-\sfrac{9}{2}\hbar^2T^6_{44}+27\hbar^2T^5_{44}-\sfrac{9}{2}\hbar^2T^4_{44}+54\hbar^2T^3_{44}\\[4pt]
	-\sfrac{27}{8}\hbar^4T^3_{22}+\sfrac{81}{8}\hbar^4T^2_{22}+\hbar^6T^0_{00}\ .
\end{multline}
The individual contributions are as follows,
\\[8pt]
{\small
\begin{equation}
\begin{aligned}
T^{9}_{66} &\=
(20|20|20|01|01|01|01|01|01)
-6(20|20|11|10|01|01|01|01|01)
+3(20|20|10|10|02|01|01|01|01)
\qquad\quad{}\\ &\ \ \,
+12(20|11|11|10|10|01|01|01|01)
-12(20|11|10|10|10|02|01|01|01)
+3(20|10|10|10|10|02|02|01|01)\\ &\ \ \,
-8(11|11|11|10|10|10|01|01|01)
+12(11|11|10|10|10|10|02|01|01)
-6(11|10|10|10|10|10|02|02|01)\\ &\ \ \,
+(10|10|10|10|10|10|02|02|02)\ ,
\end{aligned}
\end{equation}
\\[-10pt]
\begin{equation}
\begin{aligned}
T^{8}_{66} &\=
(30|30|01|01|01|01|01|01)
-6(30|21|10|01|01|01|01|01)
-6(30|20|11|01|01|01|01|01)
\qquad\qquad\qquad\quad{}\\ &\ \ \,
+6(30|20|10|02|01|01|01|01)
+6(30|12|10|10|01|01|01|01)
+12(30|11|11|10|01|01|01|01)\\ &\ \ \,
-18(30|11|10|10|02|01|01|01)
-2(30|10|10|10|03|01|01|01)
+6(30|10|10|10|02|02|01|01)\\ &\ \ \,
+9(21|21|10|10|01|01|01|01)
+6(21|20|20|01|01|01|01|01)
-6(21|20|11|10|01|01|01|01)\\ &\ \ \,
-6(21|20|10|10|02|01|01|01)
-18(21|12|10|10|10|01|01|01)
-12(21|11|11|10|10|01|01|01)\\ &\ \ \,
+30(21|11|10|10|10|02|01|01)
+6(21|10|10|10|10|03|01|01)
-12(21|10|10|10|10|02|02|01)\\ &\ \ \,
+6(20|20|20|02|01|01|01|01)
-12(20|20|12|10|01|01|01|01)
-6(20|20|11|11|01|01|01|01)\\ &\ \ \,
-24(20|20|11|10|02|01|01|01)
+6(20|20|10|10|03|01|01|01)
+12(20|20|10|10|02|02|01|01)\\ &\ \ \,
+30(20|12|11|10|10|01|01|01)
-6(20|12|10|10|10|02|01|01)
+24(20|11|11|11|10|01|01|01)\\ &\ \ \,
+12(20|11|11|10|10|02|01|01)
-18(20|11|10|10|10|03|01|01)
-24(20|11|10|10|10|02|02|01)\\ &\ \ \,
+6(20|10|10|10|10|03|02|01)
+6(20|10|10|10|10|02|02|02)
+9(12|12|10|10|10|10|01|01)\\ &\ \ \,
-12(12|11|11|10|10|10|01|01)
-6(12|11|10|10|10|10|02|01)
-6(12|10|10|10|10|10|03|01)\\ &\ \ \,
+6(12|10|10|10|10|10|02|02)
-24(11|11|11|11|10|10|01|01)
+24(11|11|11|10|10|10|02|01)\\ &\ \ \,
+12(11|11|10|10|10|10|03|01)
-6(11|11|10|10|10|10|02|02)
-6(11|10|10|10|10|10|03|02)\\ &\ \ \,
+(10|10|10|10|10|10|03|03)\ ,
\end{aligned}
\end{equation}
\\[-10pt]
\begin{equation}
\begin{aligned}
{}\;T^{7}_{66} &\=
3(30|30|02|01|01|01|01)
-6(30|21|11|01|01|01|01)
-12(30|21|10|02|01|01|01)
-2(30|20|12|01|01|01|01)\\ &\ \ \,
-10(30|20|11|02|01|01|01)
+2(30|20|10|03|01|01|01)
+10(30|20|10|02|02|01|01)
+16(30|12|11|10|01|01|01)\\ &\ \ \,
+4(30|12|10|10|02|01|01)
+8(30|11|11|11|01|01|01)
-4(30|11|11|10|02|01|01)
-10(30|11|10|10|03|01|01)\\ &\ \ \,
-6(30|11|10|10|02|02|01)
+2(30|10|10|10|03|02|01)
+2(30|10|10|10|02|02|02)
+5(21|21|20|01|01|01|01)\\ &\ \ \,
+8(21|21|11|10|01|01|01)
+14(21|21|10|10|02|01|01)
+10(21|20|20|02|01|01|01)
-14(21|20|12|10|01|01|01)\\ &\ \ \,
-4(21|20|11|11|01|01|01)
-2(21|20|11|10|02|01|01)
+4(21|20|10|10|03|01|01)
-14(21|20|10|10|02|02|01)\\ &\ \ \,
-26(21|12|11|10|10|01|01)
-14(21|12|10|10|10|02|01)
-16(21|11|11|11|10|01|01)
+32(21|11|11|10|10|02|01)\\ &\ \ \,
+16(21|11|10|10|10|03|01)
-6(21|11|10|10|10|02|02)
-2(21|10|10|10|10|03|02)
+2(20|20|20|03|01|01|01)\\ &\ \ \,
+2(20|20|20|02|02|01|01)
-6(20|20|12|11|01|01|01)
-14(20|20|12|10|02|01|01)
-4(20|20|11|11|02|01|01)\\ &\ \ \,
-6(20|20|11|10|03|01|01)
-4(20|20|11|10|02|02|01)
+10(20|20|10|10|03|02|01)
+2(20|20|10|10|02|02|02)\\ &\ \ \,
+14(20|12|12|10|10|01|01)
+32(20|12|11|11|10|01|01)
-2(20|12|11|10|10|02|01)
-12(20|12|10|10|10|03|01)\\ &\ \ \,
+10(20|12|10|10|10|02|02)
+2(20|11|11|11|11|01|01)
+8(20|11|11|11|10|02|01)
-4(20|11|11|10|10|03|01)\\ &\ \ \,
-4(20|11|11|10|10|02|02)
-10(20|11|10|10|10|03|02)
+3(20|10|10|10|10|03|03)
+8(12|12|11|10|10|10|01)\\ &\ \ \,
+5(12|12|10|10|10|10|02)
-16(12|11|11|11|10|10|01)
-4(12|11|11|10|10|10|02)
-6(12|11|10|10|10|10|03)\\ &\ \ \,
-4(11|11|11|11|11|10|01)
+8(11|11|11|10|10|10|03)
+2(11|11|11|11|10|10|02)\ ,
\end{aligned}
\end{equation}
\\[-10pt]
\begin{equation}
\begin{aligned}
T^{6}_{66} &\=
8(30|30|03|01|01|01)
+21(30|30|02|02|01|01)
-24(30|21|12|01|01|01)
-84(30|21|11|02|01|01)\\ &\ \ \, 
-24(30|21|10|03|01|01)
-42(30|21|10|02|02|01)
-24(30|20|12|02|01|01)
-30(30|20|11|03|01|01)\\ &\ \ \,
-12(30|20|11|02|02|01)
+54(30|20|10|03|02|01)
+12(30|20|10|02|02|02)
+48(30|12|12|10|01|01)\\ &\ \ \,
+96(30|12|11|11|01|01)
+24(30|12|11|10|02|01)
-24(30|12|10|10|03|01)
+30(30|12|10|10|02|02)\\ &\ \ \,
+12(30|11|11|11|02|01)
-36(30|11|11|10|03|01)
-12(30|11|11|10|02|02)
-30(30|11|10|10|03|02)\\ &\ \ \,
+8(30|10|10|10|03|03)
+16(21|21|21|01|01|01)
+66(21|21|20|02|01|01)
-24(21|21|12|10|01|01)\\ &\ \ \,
-12(21|21|11|11|01|01)
+144(21|21|11|10|02|01)
+48(21|21|10|10|03|01)
-9(21|21|10|10|02|02)\\ &\ \ \,
+30(21|20|20|03|01|01)
+12(21|20|20|02|02|01)
-54(21|20|12|11|01|01)
-138(21|20|12|10|02|01)\\ &\ \ \,
+12(21|20|11|11|02|01)
+24(21|20|11|10|03|01)
-36(21|20|11|10|02|02)
-24(21|20|10|10|03|02)\\ &\ \ \,
-24(21|12|12|10|10|01)
-132(21|12|11|11|10|01)
-54(21|12|11|10|10|02)
-24(21|12|10|10|10|03)\\ &\ \ \,
-24(21|11|11|11|11|01)
+36(21|11|11|11|10|02)
+96(21|11|11|10|10|03)
+12(20|20|20|03|02|01)\\ &\ \ \,
-2(20|20|20|02|02|02)
-9(20|20|12|12|01|01)
-36(20|20|12|11|02|01)
-42(20|20|12|10|03|01)\\ &\ \ \,
+12(20|20|12|10|02|02)
-12(20|20|11|11|03|01)
+6(20|20|11|11|02|02)
-12(20|20|11|10|03|02)\\ &\ \ \,
+21(20|20|10|10|03|03)
+144(20|12|12|11|10|01)
+66(20|12|12|10|10|02)
+36(20|12|11|11|11|01)\\ &\ \ \,
+12(20|12|11|11|10|02)
-84(20|12|11|10|10|03)
-6(20|11|11|11|11|02)
+12(20|11|11|11|10|03)\\ &\ \ \,
+16(12|12|12|10|10|10)
-12(12|12|11|11|10|10)
-24(12|11|11|11|11|10)
+2(11|11|11|11|11|11)\ ,
\end{aligned}
\end{equation}
\\[-10pt]
\begin{equation}
\begin{aligned}
T^{5}_{66} &\=
12(30|30|03|02|01)
+(30|30|02|02|02)
-36(30|21|12|02|01)
-24(30|21|11|03|01)
\qquad\qquad\qquad\quad \\ &\ \ \,
-6(30|21|11|02|02)
-12(30|21|10|03|02)
-12(30|20|12|03|01)
+6(30|20|12|02|02)\\ &\ \ \,
-6(30|20|11|03|02)
+12(30|20|10|03|03)
+48(30|12|12|11|01)
+24(30|12|12|10|02)\\ &\ \ \,
-24(30|12|11|10|03)
+4(30|11|11|11|03)
+24(21|21|21|02|01)
+24(21|21|20|03|01)\\ &\ \ \,
-3(21|21|20|02|02)
-24(21|21|12|11|01)
-12(21|21|12|10|02)
+48(21|21|11|10|03)\\ &\ \ \,
+12(21|21|11|11|02)
+6(21|20|20|03|02)
-12(21|20|12|12|01)
-36(21|20|12|10|03)\\ &\ \ \,
-6(21|20|12|11|02)
-24(21|12|12|11|10)
-12(21|12|11|11|11)
+(20|20|20|03|03)\\ &\ \ \,
-3(20|20|12|12|02)
-6(20|20|12|11|03)
+24(20|12|12|12|10)
+12(20|12|12|11|11)\ ,
\end{aligned}
\end{equation}
\\[-10pt]
\begin{equation}
\begin{aligned}
T^{4}_{66} &\=
(30|30|03|03)
-6(30|21|12|03)
+4(30|12|12|12)
+4(21|21|21|03)
-3(21|21|12|12)\ ,
\qquad\qquad\ {}
\end{aligned}
\end{equation}

\newpage

\begin{equation}
\begin{aligned}
T^{6}_{44} &\=
(20|20|01|01|01|01)
-4(20|11|10|01|01|01)
+2(20|10|10|02|01|01)\\ &\ \ \,
+4(11|11|10|10|01|01)
-4(11|10|10|10|02|01)
+(10|10|10|10|02|02)\ ,
\qquad\qquad\qquad\qquad\qquad\qquad\ {}
\end{aligned}
\end{equation}
\\[-10pt]
\begin{equation}
\begin{aligned}
T^{5}_{44} &\=
(20|20|02|01|01)
-(20|11|11|01|01)
-2(20|11|10|02|01)
+(20|10|10|02|02)
\qquad\qquad\qquad\qquad\quad\ \\ &\ \ \,
+2(11|11|11|10|01)
-(11|11|10|10|02)\ ,
\end{aligned}
\end{equation}
\\[-10pt]
\begin{equation}
\begin{aligned}
T^{4}_{44} &\=
4(17{-}6\lambda)(30|12|01|01)
-4(17{-}6\lambda)(30|11|02|01)
-4(17{-}6\lambda)(30|10|03|01)
\qquad\qquad\qquad\qquad\quad \\ &\ \ \, 
+4(17{-}6\lambda)(30|10|02|02)
-4(17{-}6\lambda)(21|21|01|01)
+4(17{-}6\lambda)(21|20|02|01)\\ &\ \ \,
+4(17{-}6\lambda)(21|12|10|01)
+8(17{-}6\lambda)(21|11|11|01)
-12(17{-}6\lambda)(21|11|10|02)\\ &\ \ \,
+4(17{-}6\lambda)(21|10|10|03)
+4(17{-}6\lambda)(20|20|03|01)
-(59-24\lambda)(20|20|02|02)\\ &\ \ \,
-12(17{-}6\lambda)(20|12|11|01)
+4(17{-}6\lambda)(20|12|10|02)
+2(59{-}24\lambda)(20|11|11|02)\\ &\ \ \,
-4(17{-}6\lambda)(20|11|10|03)
-4(17{-}6\lambda)(12|12|10|10)
+8(17{-}6\lambda)(12|11|11|10)\\ &\ \ \,
-(59-24\lambda)(11|11|11|11)\ ,
\end{aligned}
\end{equation}
\\[-10pt]
\begin{equation}
\begin{aligned}
T^{3}_{44} &\=
(17{-}6\lambda)(30|12|02)
-(17{-}6\lambda)(30|11|03)
-(17{-}6\lambda)(21|21|02) \\ &\ \ \,
+(17{-}6\lambda)(21|20|03)
+(17{-}6\lambda)(21|12|11)
-(17{-}6\lambda)(20|12|12)\ ,
\qquad\qquad\qquad\qquad\qquad\qquad\quad\ {}
\end{aligned}
\end{equation}
\\[-10pt]
\begin{equation}
\begin{aligned}
T^{3}_{22} &\=
\big(177-16\lambda(13{-}3\lambda)\bigr)(20|01|01)
-2\big(177-16\lambda(13{-}3\lambda)\bigr)(11|10|01) 
\qquad\qquad\qquad\qquad\qquad\qquad\  \\ &\ \ \,
+\big(177-16\lambda(13{-}3\lambda)\bigr)(10|10|02)\ ,
\end{aligned}
\end{equation}
\\[-10pt]
\begin{equation}
\begin{aligned}
T^{2}_{22} &\=
\big(177-16\lambda(13{-}3\lambda)\bigr)(20|02)
-\big(177-16\lambda(13{-}3\lambda)\bigr)(11|11)\ ,
\qquad\qquad\qquad\qquad\qquad\qquad\quad\ \, {}
\end{aligned}
\end{equation}
\\[-10pt]
\begin{equation}
\begin{aligned}
T^{0}_{00} &\=
\sfrac{17001}{32} -\sfrac{81}{4}\lambda(43{-}6\lambda)\ ,
\qquad\qquad\qquad\qquad\qquad\qquad\qquad\qquad
\qquad\qquad\qquad\qquad\qquad\qquad\qquad{}
\end{aligned}
\end{equation}
}
\\[8pt]
where the constant last term is of course irrelevant.

\newpage
\noindent {\bf Acknowledgements}

\noindent
FC was supported by Fondecyt grants 1171475 and 1211356. 
He thanks the Departamento de F\'isica Te\'orica, At\'omica y \'Optica at Universidad de Valladolid
and Universidad Austral de Chile,
where this project was initiated, for its kind hospitality.


\section*{}

\end{document}